\documentclass[twocolumn,showpacs,preprintnumbers,showkeys,superscriptaddress]{revtex4}
\usepackage{graphicx}
\usepackage{dcolumn}
\usepackage{bm}

\def\eqref#1{Eq.~(\ref{eq:#1})}
\def\eqlab#1{\label{eq:#1}}

\newcommand{\vslash}[1]{#1\hspace{-0.5em}/}

\begin{document}

\title{Comparing Phenomenological recipes with a Microscopic model for the
Electric Amplitude in Strangeness Photoproduction}
\author{A.Yu.\ Korchin}     \email{korchin@kvi.nl}
\affiliation{Kernfysisch Versneller Instituut, University of Groningen,
9747 AA Groningen, The~Netherlands}
\affiliation{NSC `Kharkov Institute of Physics and Technology', 61108 Kharkov, Ukraine}
\author{O. Scholten}     \email{scholten@kvi.nl}
\homepage{http://www.kvi.nl/~scholten}
\affiliation{Kernfysisch Versneller Instituut, University of Groningen,
9747 AA Groningen, The~Netherlands}
\date{today}

\begin{abstract}
Corrections to the Born approximation in photo-induced strangeness production
off a proton are calculated in a semi-realistic microscopic model. The vertex
corrections and internal contributions to the amplitude of the $\gamma p \to
K^+ \Lambda$ reaction are included on the one-loop level. Different
gauge-invariant phenomenological prescriptions for the modification of the Born
contribution via the introduction of form factors and contact terms are
discussed. In particular, it is shown that the popular minimal-substitution
method of Ohta corresponds to a special limit of the more realistic approach.
\end{abstract}

\maketitle

\section{\label{sec:Introduction}Introduction}

One of the goals in studies of photo- and electroproduction of pions or kaons
on the proton is the extraction of information on baryon
resonances~\cite{Mar95}. This is usually done on the basis of models built on
effective Lagrangians~\cite{Feu99,Pen02,Jan02,Kon01}. A typical amplitude,
describing meson photoproduction, includes the Born contribution as a kind of
background to be added to the resonance contributions. Only at low photon
energies the physics is determined by the Born diagrams. With increasing photon
energy the Born contribution rises to produce unrealistically large cross
sections at energies of interest for strangeness production as noticed by many
authors~\cite{Wor92,Hab98,Dav01}. The popular strategy is to suppress the Born
amplitude by including form factors (FF's) in the strong-interaction vertices.
These FF's account for physics on the scales beyond what is considered,
\textit{i.e.} exchanges of heavier mesons which are truncated from the
Lagrangian and (higher order) loop corrections which are omitted for
simplicity. Closely associated with FF's are additional terms, called contact
terms (sometimes 4-point vertices or internal contributions), which restore
gauge invariance (GI) usually violated by the introduction of FF's. Several
different prescriptions for the FF's and contact terms are commonly in use.
While at low energies these prescriptions lead to relatively close results, at
higher energies the calculated cross sections may differ drastically. Therefore
the information on the properties of resonances extracted from these processes
is strongly influenced by uncertainties in the treatment of this
problem~\cite{Jan02}.

In the present work we study two commonly used procedures: Ohta's
minimal-substitution method~\cite{Oht89} and Davidson -
Workman's (DW) recipe~\cite{Dav01,Dav01-2}. The most essential difference in
predictions of these two methods for the meson photoproduction processes, such
as $\gamma N\rightarrow \pi N$ or $\gamma p\rightarrow K^{+}\Lambda $ ($K\Sigma
$), is the modification of one particular invariant amplitude, $A_{2}(s,t)$.
This amplitude is related to the electric part of the amplitude and originates
from the convection current of the charged particles involved in the reaction.
Its contribution to the matrix element, being proportional to the momenta of
particles, affects the cross section at high energies. In the approach of Ohta
$A_{2}(s,t)$ is not altered, due to the complete cancellation between the
effects of FF's and the contact terms. In other methods, in particular in
DW~\cite{Dav01,Dav01-2}, or Haberzettl's approach~\cite{Hab97,Hab98}, the
amplitude $A_{2}(s,t)$ changes considerably compared to the Born amplitude.

In order to study different phenomenological approaches we calculate vertex
corrections and internal contributions to the electric amplitude in an
effective Lagrangian model. Irreducible one-loop contributions are included for
the reaction $\gamma p\rightarrow K^{+}\Lambda$. This allows us to extract,
within a gauge-invariant model, both the FF's (associated with 3-point loop
corrections) and the contact terms (associated with 4-point loop corrections),
and to make a comparison with phenomenological approaches. The model is $
SU(3)_{flavor}$ symmetrical and describes the baryon-meson interaction as well
as the meson-meson interaction of the scalar and pseudo-scalar mesons. In the
intermediate states of the diagrams we include the scalar meson, kaon, proton
and lambda-hyperon. The pion and sigma-hyperon are not included as yet, and
thus the model can be considered as semi-realistic. On the one-loop level there
appear three diagrams for the $K^{+}p\Lambda$ vertex and four diagrams for the
internal amplitude. The vertex corrections have the property that in the limit
of large mass of the scalar meson each of the three vertex-correction diagrams
generates FFs, which depend exclusively on one of the Mandelstam variables
$s,u$ or $t$. This in turn leads to an interesting effect of
cancellation between the vertex corrections and the 4-point diagrams
in the scalar amplitude $A_{2}(s,t)$.

We should mention that some loop contributions in the pion photoproduction on
the nucleon were studied in Ref.~\cite{Bos92}, where the need for consistent
treatment of corrections to the Born amplitude was stressed.

The paper is organized as follows. In Sect.~\ref{subsec:Born} the Born
approximation is briefly discussed and invariant amplitudes are introduced. The
structure of $K^{+} p \Lambda $ vertex is addressed in sect.~\ref
{subsec:Vertex}. Different recipes for restoring GI are outlined. The present
model for calculation of the loop corrections and the parameters are described
in sect.~\ref{subsec:Loops}. Results of calculations and a discussion are
presented in sect.~\ref{sec:Results}. In sect.~\ref {sec:Conclusions} we draw
conclusions. Finally, Appendix~\ref{App:A} contains details of the calculation
of the 3- and 4-point loop integrals.


\section{\label{sec:Formalism} Formalism}

\subsection{\label{subsec:Born} Born diagrams}

\begin{figure}[tbp]
\includegraphics[width=8cm]{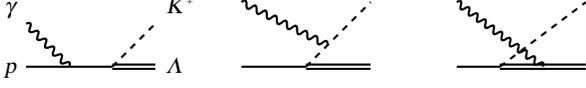}
\caption[Born]{Born diagrams for the reaction $\protect\gamma +p \to K^{+} +
\Lambda$. Solid lines depict the proton, double-solid lines - the lambda
hyperon, wavy lines - the photon, and dashed lines - the kaon.}
\label{fig:1}
\end{figure}

The Born amplitude for the reaction $\gamma +p\longrightarrow K^{+}+\Lambda $
 (see Fig.~\ref{fig:1}) can be split in electric and magnetic parts, {\it i.e.}\ $
\mathcal{T}_{B}^{\mu }=\mathcal{T}_{B,El}^{\mu}+\mathcal{T}_{B,Mag}^{\mu}$, where
\begin{eqnarray}
\mathcal{T}_{B,El}^{\mu } &=&eg\bar{u}(p^{\prime })\gamma _{5}u(p)\left(
\frac{2p^{\mu }}{s-M_{N}^{2}}+\frac{2q^{\mu }}{t-\mu _{k}^{2}}\right) ,
\nonumber \\
\mathcal{T}_{B,Mag}^{\mu } &=&eg\bar{u}(p^\prime ) \gamma_{5} \Big\{
 \frac{1}{s-M_{N}^{2}} \big[ (1+\kappa_p) \gamma^\mu  \vslash{k}+ \nonumber \\
 &&\frac{\kappa_p}{M_N} (\gamma^\mu k\cdot p - p^\mu \vslash{k}) \big]
 +\frac{1}{u-M_\Lambda^2} \big[-\kappa_\Lambda \gamma^\mu \vslash{k}
 \nonumber \\
 &&+\frac{\kappa_\Lambda}{M_\Lambda}
 (\gamma^\mu k\cdot p^\prime -p^{\prime \mu} \vslash{k})
 \big] \Big\} u(p)\;.  \label{eq:1}
\end{eqnarray}
The 4-momenta of the (real) photon, proton, kaon and lambda are denoted by $
k,p,q$ and $p^{\prime }$ respectively, and $M_{N},\mu_k$ and $M_\Lambda $ are
the masses of the proton, kaon and lambda. The invariants $s,t,u$ are the
Mandelstam variables satisfying the relation $s+t+u=M_N^2+M_\Lambda^{2}+
\mu_k^2$. The anomalous magnetic moments of the proton and lambda are
denoted by $\kappa_{p}$ and $\kappa_{\Lambda }$ respectively. Finally, $g$
stands for the $K^{+}p\Lambda $ coupling constant. The spinor of the initial
proton is $u(p)$, and the one of the final lambda is $\bar{u}(p^{\prime })$
where the helicity (spin) indices are
suppressed. The amplitudes in Eqs.(\ref{eq:1}) are gauge invariant, {\it i.e.}\ $%
k\cdot \mathcal{T}_{El}=k\cdot \mathcal{T}_{Mag}=0.$

We will use the formalism~\cite{Bal61,Bar68} in which the amplitude is
decomposed through the independent spin-tensors
\begin{equation}
\mathcal{T}^{\mu }=\sum_{i=1}^{4} A_{i}(s,t)\bar{u}(p^{\prime })
\mathcal{M}_{i}^{\mu }u(p)\;,  \label{eq:2}
\end{equation}
with
\begin{eqnarray}
\begin{array}{l}
\mathcal{M}_{1}^{\mu }=-\gamma _{5}\gamma ^{\mu }\vslash{k}\;, \\
 \mathcal{M}_{2}^{\mu }=2\gamma _{5}(p^\mu k\cdot
p^{\prime }-p^{\prime \mu }k\cdot p)\;, \\
 \mathcal{M}_{3}^{\mu }=\gamma_{5}(\gamma^\mu k\cdot p - p^\mu \vslash{k})\;,\\
 \mathcal{M}_{4}^{\mu }=\gamma _{5}(\gamma ^{\mu }k\cdot
p^{\prime}-p^{\prime\mu }\vslash{k})\;.
\end{array}
\label{eq:3}
\end{eqnarray}
The gauge invariant basis $\mathcal{M}_{i}^{\mu }$ is constructed in such way that the
scalar amplitudes $A_{i}(s,t)$ are free from kinematical singularities and
zeros~\cite{Bar68}. For the Born diagrams $A_{i}$ take the form (we omit
arguments for brevity)
\begin{eqnarray}
\begin{array}{l}
 A_{1}^{Born}=eg\left( \frac{1+\kappa _{p}}{s-M_{N}^{2}}+
 \frac{\kappa_{\Lambda }}{u-M_{\Lambda }^{2}}\right) \;,\\
 A_{2}^{Born}=eg\frac{2}{(s-M_{N}^{2})(t-\mu _{K}^{2})}\;, \\
 A_{3}^{Born}=eg\frac{\kappa _{p}}{M_{N}}\frac{1}{s-M_{N}^{2}}\;, \\
 A_{4}^{Born}=eg\frac{\kappa _{\Lambda }}{M_{\Lambda }} \frac{1}{u-M_{\Lambda}^{2}}\;.
\end{array}
\label{eq:4}
\end{eqnarray}
The magnetic part results in the single-pole amplitudes $A_{1}^{Born},
A_{3}^{Born}$ and $A_{4}^{Born}$, while the electric part contributes solely to
the double-pole amplitude $A_{2}^{Born}$. As was discussed in \cite{Bar68} the
latter is a peculiar feature of the amplitude with real photons and the choice
of the spin-tensors $\mathcal{M}_{i}^{\mu }$.

\subsection{\label{subsec:Vertex} $K^{+}p \Lambda $ vertex and form factors}

In a phenomenological description strong FF's are often included directly in
the Born diagrams. At this point we recall the general structure of the
$K^{+}p\Lambda $ vertex
\begin{eqnarray}
\Gamma (p^{\prime },p;q)&=&
 g\Bigg( \gamma_{5} f_{1}+ \gamma_{5} \frac{\vslash{p}-M_{N}} {M_{N}}
 f_{2}+ \frac{\vslash{p}^{\prime} -M_{\Lambda}} {M_{\Lambda}}
 \gamma_{5} f_{3} \nonumber \\
 &+& \frac{\vslash{p}^{\prime }-M_{\Lambda}}{M_{\Lambda }}
 \gamma_{5} \frac{\vslash{p}-M_{N}}{M_{N}} f_{4} \Bigg)\;,
\label{eq:5}
\end{eqnarray}
where $p^{\prime}, p$ and $q$ are the lambda, proton and kaon momenta
respectively, and $f_{i}\equiv f_{i}(p^{\prime 2},p^{2},q^{2})$ are scalar
functions. If only one of the hadrons is off its mass shell then
Eq.(\ref{eq:5}) simplifies. For this situation it is convenient to introduce
the 3-point FF's:
\begin{eqnarray}
\begin{array}{l}
 F_{s}(p^{2})=f_{1}(M_{\Lambda }^{2},p^{2},\mu _{K}^{2})\;, \\
 F_{u}(p^{\prime 2})=f_{1}(p^{\prime 2},M_{N}^{2},\mu_{K}^{2})\;, \\
 F_{t}(q^{2})=f_{1}(M_{\Lambda }^{2},M_{N}^{2},q^{2})\;,   \\
 G_{s}(p^{2})=f_{2}(M_{\Lambda }^{2},p^{2},\mu _{K}^{2})\;, \\
 G_{u}(p^{\prime2})=f_{3}(p^{\prime 2},M_{N}^{2},\mu _{K}^{2})\;.
\end{array}
\label{eq:6}
\end{eqnarray}
In general the functions $F_{s}(p^{2}),$ $F_{u}(p^{\prime 2})$ and
$F_{t}(q^{2})$ have different functional dependencies as indicated by the
subscript $s$, $u$ or $t,$ and are normalized to unity on the mass shell.

When the vertex in Eq.~(\ref{eq:5}) is included in the tree-level terms, the
magnetic amplitude $\mathcal{T}_{B,Mag}^{\mu }$ in Eq.~(\ref{eq:1}) is modified
to $\mathcal{T}_{Mag}^{\mu }$ with the following result for the scalar
amplitudes
\begin{eqnarray}
A_{1,Mag} &=&eg\Big( \frac{1+\kappa_p}{s-M_N^2} F_s(s)
 + \frac{\kappa_\Lambda}{u-M_\Lambda^2} F_u(u) \nonumber \\
 &+&G_s(s) \frac{\kappa_p}{2M_N^2}
 + G_u(u) \frac{\kappa_\Lambda}{2M_\Lambda^2} \Big) \;,
\nonumber \\
A_{3,Mag} &=&eg\left( \frac{\kappa _{p}}{M_{N}}\frac{1}{s-M_{N}^{2}}F_{s}(s)+%
\frac{2}{M_{N}}\frac{1}{s-M_{N}^{2}}G_{s}(s)\right)\;,  \nonumber \\
A_{4,Mag} &=&eg\frac{\kappa _{\Lambda }}{M_{\Lambda }}\frac{1}{u-M_{\Lambda
}^{2}}F_{u}(u) \;.
\label{eq:7}
\end{eqnarray}
The electric amplitude changes to
\begin{eqnarray}
\mathcal{T}_{El}^{\mu }&=&
 eg\bar{u}(p^{\prime }) \gamma_5 \Big( \frac{2p^\mu}{s-M_N^2}
 \big[ F_{s}(s)+G_{s}(s)\frac{\vslash{k}}{M_N} \big] \nonumber \\
 &+& \frac{2q^\mu}{t-\mu_k^2} F_{t}(t) \Big) u(p)\;.  \label{eq:8}
\end{eqnarray}
This term cannot be cast in the form of Eq.~(\ref{eq:2}) since it is not gauge
invariant. Indeed, contraction with the photon momentum results in
\begin{eqnarray}
&&k\cdot \mathcal{T}_{El} = eg\bar{u}(p^{\prime })\gamma_{5}
 \big[ F_{s}(s)+G_{s}(s)\frac{\vslash{k}}{M_{N}}-F_{t}(t) \big]u(p)
\nonumber \\
&& = e\bar{u}(p^\prime) \big[ \Gamma(p^\prime,p+k;q) -
 \Gamma(p^\prime,p;q-k) \big] u(p)\neq 0 \;.
\label{eq:9}
\end{eqnarray}
In general, it is known that there are other contributions to the amplitude
\cite{Gel54,Kaz59,Bos92} which ensure GI of the total amplitude. We will denote
this additional amplitude by $\mathcal{T}_{c}^{\mu }$ and discuss
different ways of constructing this amplitude.

\subsubsection{Minimal-substitution method of Ohta~\cite{Oht89} }

In the original formulation of Ohta~\cite{Oht89} the electromagnetic
interaction (EM) was included directly in the 3-point $\pi NN$ \ vertex using
the minimal-substitution method. This allowed for construction of
$\mathcal{T}_{c}^{\mu}$ in terms of the FF's. It may be instructive to derive
the same result (for the $pK^{+}\Lambda$ vertex) using a simpler, though less
rigorous, method which was applied in Ref.~\cite{Sch96}. The GI requirement for
the total amplitude $\mathcal{T}_{Mag}+\mathcal{T}_{El}+\mathcal{T}_{c}$, with
the help of Eq.~(\ref{eq:9}) and definitions \ $s-M_{N}^{2}= 2k\cdot p$, $\ t-\mu
_{k}^{2}=-2k\cdot q$, can be written as
\begin{eqnarray}
k\cdot \mathcal{T}_{c} &=& -k\cdot \mathcal{T}_{El}=
 egk_\mu \bar{u}(p^\prime) \gamma_5 \Big( 2p^\mu
 \frac{1-F_s(s)}{s-M_N^2} \nonumber \\
 &+& 2q^\mu \frac{1-F_t(t)}{t-\mu_K^2}
 -G_{s}(s)\frac{\gamma^\mu}{M_N} \Big) u(p)\;,
\label{eq:10}
\end{eqnarray}
from which $\mathcal{T}_{c}^{\mu}$ is obtained as the term multiplying $k_{\mu}$.
We now find (using \eqref{8})
\begin{eqnarray}
\mathcal{T}_{El}^{\mu } &+& \mathcal{T}_{c}^\mu = eg\bar{u}(p^{\prime })
 \gamma_5 \Big[ \frac{2p^\mu}{s-M_N^2}+\frac{2q^\mu}{t-\mu_K^2}
 \nonumber \\
 &+& G_{s}(s)\frac{1}{M_{N}}\left( \frac{p^{\mu }}{k\cdot p}
\vslash{k} -\gamma ^{\mu }\right) \Big] u(p)  \label{eq:11}
\end{eqnarray}
up to the transverse terms which are not constrained by the condition of GI.
The amplitude in Eq.~(\ref{eq:11}) is apparently gauge invariant and yields two
scalar amplitudes
\begin{eqnarray}
 A_{2,El}=eg\frac{2}{(s-M_{N}^{2})(t-\mu _{K}^{2})} \;, \nonumber \\
 A_{3,El}=-eg\frac{1}{M_{N}}\frac{2}{s-M_{N}^{2}}G_{s}(s) \;.
 \label{eq:12}
\end{eqnarray}
It is seen that, firstly, $A_{3,El}$ cancels the term in $A_{3,Mag}$
proportional to $G_{s}(s)$, and secondly, the amplitude $A_{2,El}$ coincides
with the Born amplitude $A_{2}^{Born}$ in Eq.~(\ref{eq:4}). The main result is
thus that the scalar amplitude $A_{2}$ is not modified in the presence of
strong FF's, as noticed earlier in~\cite{Wor92}.

\subsubsection{\label{subsec:DW} Approach of Davidson and Workman~\cite{Dav01} }

In order to change the electric contribution in a phenomenological approach
some authors introduced FF's directly in the amplitude $A_{2}$ in
Eq.~(\ref{eq:2}). As was pointed out in~\cite{Dav01,Dav01-2}, care should be
taken with the structure of these FF's in order to avoid spurious pole
contributions as generated with the original introduction of these FF's
in~\cite{Hab97,Hab98}. The procedure in \cite{Dav01,Dav01-2}, in which the FF's
modify the total amplitude $A_{2}$, will be referred to as the DW approach.

In the DW approach the amplitude $A_{2}$ is modified to
\begin{eqnarray}
A_{2} &=&A_{2}^{Born}\widehat{F}=A_{2}^{Born}
 +eg\frac{2}{(s-M_N^2)(t-\mu_K^2)} (\widehat{F}-1) \nonumber \\
 &\equiv& A_{2}^{Born}+\Delta A_{2}^{DW}\;,
\label{eq:13}
\end{eqnarray}
where the factor $\widehat{F}$ for the reaction $\gamma p \to K^{+} \Lambda$ is
chosen to be
\begin{equation}
\widehat{F}=F_{s}(s)+F_{t}(t)-F_{s}(s)F_{t}(t)\;,  \label{eq:14}
\end{equation}
to ensure that the correction to the Born contribution is free from poles at
$s=M_{N}^{2}$ and $t=\mu _{K}^{2}$. The functions $F_{s}(s)$ and $F_{t}(t)$ are
normalized to unity on-shell and are usually parametrized in the monopole or
dipole form but are not necessarily related to the FF's introduced in
sect.~\ref{subsec:Vertex}.

\subsection{\label{subsec:Loops} Loop contributions}

\begin{figure}[tbp]
\includegraphics[width=8cm]{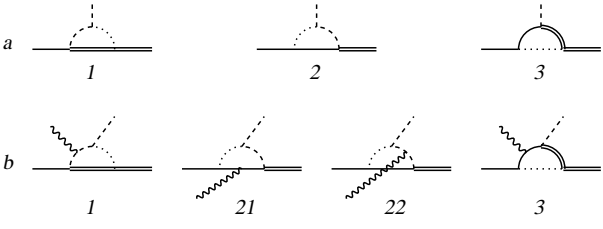}
\caption[Vertex]{One-loop corrections in the present model: $K^{+} p
\Lambda $ vertex (a),  4-point contribution
$\mathcal{L}^{\protect\mu }_{[4]}$ (b). Dotted lines correspond to the
$\protect\sigma $- meson, other notations are the same as in Fig.~\ref{fig:1}.}
\label{fig:2}
\end{figure}

In a microscopic model for the reaction mechanism there are various loop
corrections to the Born diagrams. The simplest loop corrections are
self-energy insertions in the propagators. These corrections are partially
compensated by the 3-point loop corrections to the EM vertices
$\gamma pp$ and $\gamma KK$. The net result is that only the magnetic
contribution is affected, however the convection current, which is of our main
concern, remains unchanged. For this reason these loop contributions will not
be included. We will come back to this issue in the end of
sect.~\ref{subsec:Cancellation}.
The second type of loop diagrams are corrections to
the $K^{+}p\Lambda $ vertex shown in Fig.~\ref{fig:2}a.
These are ordered in a particular way to allow for an interpretation
in terms of the 3-point FF's introduced in sect.~\ref{subsec:Vertex}.
In this section we will explicitly calculate the 3-point loop corrections,
$\mathcal{L}_{[3]}^\mu$, to the electric amplitude.
As discussed in sect.~\ref{subsec:Vertex}, GI of the
full amplitude is restored after inclusion of the 4-point diagrams (internal
amplitude) depicted in Fig.~\ref{fig:2}b. These diagrams can be obtained by
attaching the photon to the lines of charged intermediate particles in the
3-point loops and will be referred to as $\mathcal{L}_{[4]}^{\mu }$.

The amplitude $\mathcal{L}_{[4]}^{\mu }$ cannot be expanded in the basis
\eqref{2}, since only the sum $\mathcal{L}_{[3]}^\mu+ \mathcal{L}_{[4]}^{\mu}$
is gauge invariant.
To extract the amplitude $A_{2}$ we express the loop corrections in terms of
the Lorentz structures appearing in \eqref{2}:
$\gamma_5p^\mu, \gamma_5q^\mu,
\gamma_5\gamma^\mu, \gamma_5p^\mu \vslash{k},
\gamma_5q^\mu \vslash{k}, \gamma_5\gamma^\mu
\vslash{k}$, where it can be noted that
$\mathcal{M}_2^\mu=2 \gamma_5 (q^\mu k\cdot p-
p^\mu k\cdot q)$ and $\mathcal{M}_4^\mu=\mathcal{M}_3^\mu+\gamma_5 (q^\mu
\vslash{k} - \gamma^\mu k\cdot q)$.
Since $\mathcal{M}_{2}^{\mu}$ is expressed solely in terms of
$\gamma_5p^\mu$ and $\gamma_5q^\mu$, it will be sufficient to retain only
these terms from the loop corrections and write
\begin{eqnarray}
\mathcal{L}_{[3]}^{\mu} &=&
 2eg\bar{u}(p^{\prime}) \gamma_5 \Big( p^\mu \frac{F_s(s)-1}{s-M_N^2}
 \nonumber \\
 &+& q^\mu \frac{F_{t}(t)-1}{t-\mu_k^2} \Big) u(p)+...\;,
\eqlab{15a} \\
 \mathcal{L}_{[4]}^{\mu} &=&
 2eg\bar{u}(p^{\prime })\gamma_{5} \Big[ p^\mu H_{s}(s,t) \nonumber \\
 &+& q^{\mu} H_{t}(s,t) \Big] u(p)+...\;,
\label{eq:15b}
\end{eqnarray}
where the ellipsis  $ ... $ means the omitted terms containing the Lorentz
tensors $\mathcal{M}_{1}^{\mu },
\mathcal{M}_{3}^{\mu}$ and $\mathcal{M}_{4}^{\mu }$ appearing in the magnetic terms.
It should be noted that the functions $F_{t}(t)$ and $F_{s}(s)$ in the loop
correction $\mathcal{L}_{[3]}^\mu$, \eqref{15a}, coincide with the
phenomenological FF's appearing in \eqref{8}.
The condition of GI, $k \cdot \big( \mathcal{L}_{[3]} + \mathcal{L}_{[4]} \big)=0$,
imposes the relation
\begin{equation}
(s-M_{N}^{2})H_{s}(s,t) -(t-\mu _{K}^{2})H_{t}(s,t) =F_{t}(t)-F_{s}(s) \;.
\label{eq:16}
\end{equation}
The correction to the Born amplitude $A_{2}^{Born}$ can now be expressed as
\begin{eqnarray}
\Delta A_{2} &=&
 eg\frac{2}{s-M_N^2} \left( \frac{F_{t}(t)-1}{t-\mu_K^2} + H_{t}(s,t)\right)
\nonumber \\
 &=& eg\frac{2}{t-\mu_K^2} \left( \frac{F_{s}(s)-1}{s-M_N^2} +H_{s}(s,t) \right)
 \;. \label{eq:17}
\end{eqnarray}
To find the total correction to the Born amplitude one thus needs the
coefficients multiplying $q^{\mu}$ in $\mathcal{L}_{[3]}^{\mu}$ (see
\eqref{15a}) and in $\mathcal{L}_{[4]}^{\mu}$ (see
\eqref{15b}) (alternatively $\Delta A_{2}$ can be expressed
through $F_{s}(s)$ and $H_{s}(s,t)$). It should be noted that $\Delta A_{2}$ cannot
have poles, and the following condition
\begin{equation}
\lim_{s\rightarrow M_{N}^{2}}H_{t}(s,t)= \frac{1-F_{t}(t)}{t-\mu_{K}^{2}}
\label{eq:l8}
\end{equation}
should hold at $s=M_{N}^{2}$, which is the un-physical point for the
$s$-channel.

To calculate loop corrections we use an effective-Lagrangian model which is
$SU(3)_{flavor}$ symmetrical and includes as degrees of freedom the baryon
octet matrix $B$, the scalar ($J^{P}=1^{+})$ meson nonet $\Phi_{s}$, and the
pseudo-scalar ($J^{P}=1^{-})$ meson nonet $\Phi _{ps}$. The corresponding
Lagrangian is
\begin{eqnarray}
L &=& L_{0}+D_{s}tr(\Phi_{s}\{B,\bar{B}\}) +F_{s}tr(\Phi_{s}[B,\bar{B}])
\nonumber \\
 &&\hspace{-3em}+ D_{ps}tr(\Phi_{ps}\{B,\bar{B}\})+F_{ps}tr(\Phi_{ps}[B,\bar{B}])+L_{\Phi}
\;,  \label{eq:19}
\end{eqnarray}
where $L_{0}$ is the free part, $\bar{B}=B^{\dagger }\gamma _{0}$, \ $[B,%
\bar{B}]$ ($\{B,\bar{B}\}$) stands for commutator (anti-commutator), $D_{s}$
and $F_{s}$ ($D_{ps}$ and $F_{ps}$) are the baryon-meson coupling constants for
scalars~\cite{Mae89} (pseudo-scalars), and $L_{\Phi }$ is the Lagrangian
describing meson-meson interaction. The latter is chosen in the $U(3)\times
U(3)$ linear $\sigma$-model~\cite{Lev67} (the explicit form of $L_{\Phi }$\ can
be found,
e.g., in~\cite{Tor99}). For the purpose of our paper it is essential that $%
L_{\Phi }$ describes the $\sigma K^{+}K^{-}$ and $ f_{0}K^{+}K^{-}$
couplings, where $\sigma \equiv f_{0}(400-1200)$ and $f_{0}\equiv f_{0}(980)$
(or $f_{0}(1370)$) represent the scalar mesons~\cite{PDG}.

In the calculation of loops we do not include the $\pi$-meson and the
$\Sigma$-hyperon in the intermediate states. This restricts the one-loop
diagrams to those shown on Fig.~\ref{fig:2}. Calculation of the corresponding
integrals is tedious and we refer to Appendix~\ref{App:A} for details.

The coupling constants of the $SU(3)$ singlet $\phi_{0}$ and octet $\phi_{8}$
states to proton and lambda follow from Eq.~(\ref{eq:19}). To get couplings of
the physical mesons, $\sigma $ and $f_{0}$, one needs in addition the mixing
angle for the scalar mesons. All parameters are given in the Nijmegen
baryon-baryon one-boson-exchange model of Ref.~\cite{Mae89},
\begin{eqnarray}
\begin{array}{ll}
g_{\sigma pp} = 16.90\;, \; & g_{\sigma \Lambda \Lambda }=9.84\;,\\
 g_{K^{+}p \Lambda} = -14.113 \;, & g_{f_{0}pp}=-2.97\;, \\
  g_{f_{0}\Lambda \Lambda }=-9.10\;,
\end{array}
\label{eq:20}
\end{eqnarray}
and  masses $m_{\sigma }$ = 0.76 GeV, $m_{f_{0}}$ = 0.993 GeV. The
corresponding vertex is $-ig_{\sigma BB}$ ($-ig_{f_{0}BB}$).  The $\sigma
K^{+}K^{-}$ vertex is $-ig_{\sigma K^{+}K^{-}},$ with the coupling
constant~\cite{Tor99}: $g_{\sigma K^{+}K^{-}}=\sqrt{3}(m_{\sigma }^{2}-\mu
_{K}^{2})/(2f_{K})\sin (\alpha -\Delta \alpha )$, where $f_{K}=113$ MeV is the
kaon weak-decay constant, $\alpha =\arcsin (1/\sqrt{3})\approx 35.26^{\circ }$
is the ``ideal'' mixing angle, and $\Delta \alpha $ is a correction, usually of
the order $3^{\circ }-10^{\circ }$.
Likewise, the $f_{0}K^{+}K^{-}$ vertex is $-ig_{f_{0}K^{+}K^{-}}$ with $%
g_{f_{0}K^{+}K^{-}}=\sqrt{3}(m_{f_{0}}^{2}-\mu _{K}^{2})/(2f_{K})\cos
(\alpha -\Delta \alpha )$.

\section{\label{sec:Results}Results of calculation and discussion}

\subsection{\label{subsec:Results FFs}Form factors and amplitude $A_2(s,t)$ }

\begin{figure}[tbp]
\includegraphics[width=8cm,bb=58 55 552 744]{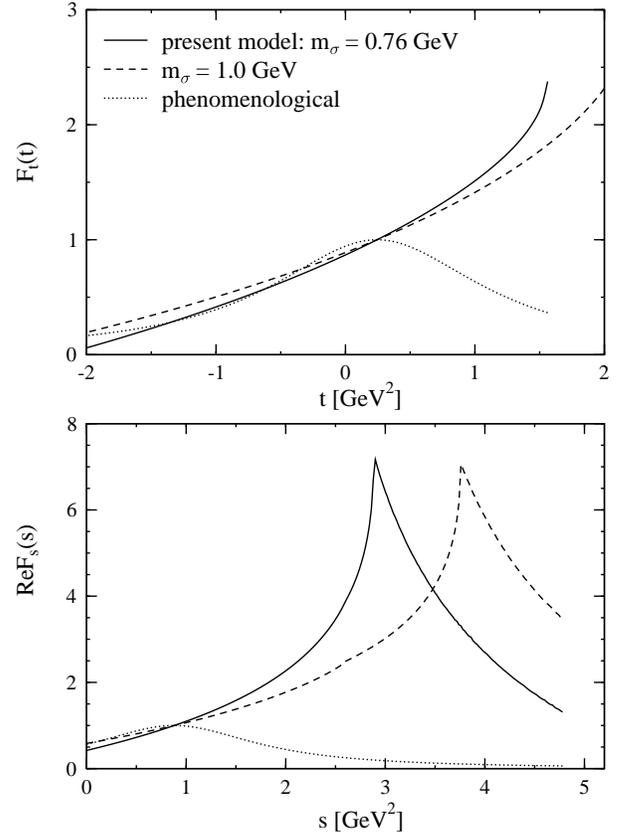}
\caption[Formfactors]{Form factors in the $t$-channel (upper panel) and $s$%
-channel (lower panel). Solid and dashed lines - present model with
$m_{\protect\sigma}$ = 0.76 GeV and 1.0 GeV respectively, dotted lines -
phenomenological form factors from Eqs.~(\ref{eq:21}).}
\label{fig:3}
\end{figure}

In Fig.~\ref{fig:3} we show the FF's as extracted from the loop corrections in
the $t$- and $s$-channels for two masses of the $\sigma $-meson, 0.76 GeV and
1.0 GeV. For comparison the phenomenological FF's used in most analyses are
also plotted. These have the typical bell-shape form (see, for
example,~\cite{Dav01})
\begin{eqnarray}
F_{s}^{(ph)}(s) &=& \frac{\Lambda^{4}}{(s-M_N^2)^2+\Lambda^{4}} \;,
 \nonumber \\
F_{t}^{(ph)}(t) &=& \frac{\Lambda^{4}}{(t-\mu_K^2)^2 +\Lambda^{4}}
\label{eq:21}
\end{eqnarray}
with a cut-off mass $\Lambda $ of about 1 GeV. All FF's are normalized to unity
at the corresponding on-shell points. We do not present FF in the $u$-channel
as it is not relevant for the discussion of the electric amplitude.

\begin{figure}[tbp]
\includegraphics[width=8cm,bb=45 60 553 744]{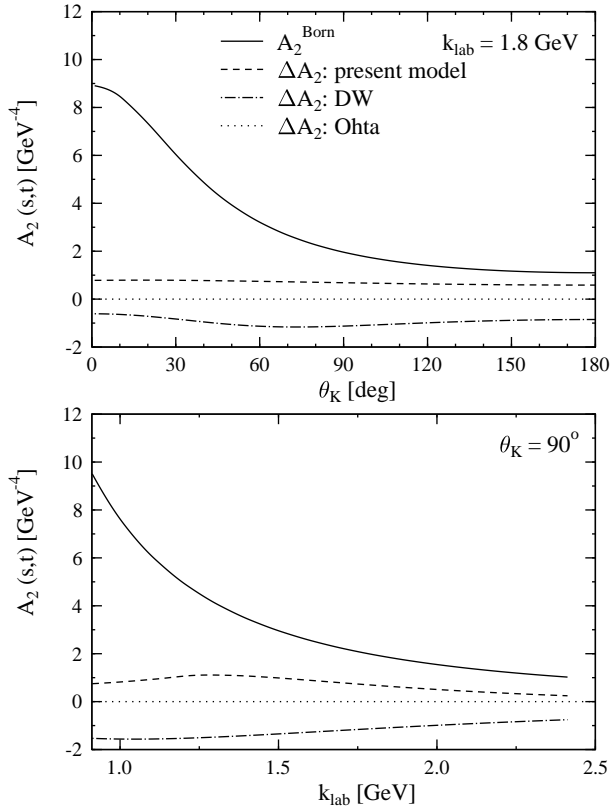}
\caption[A2]{Invariant amplitude $A_2(s,t)$ (real part) as function
of scattering angle at fixed photon lab energy (top), and as
function of energy at fixed angle (bottom).
The calculation is performed with $m_{\protect\sigma}$ = 0.76 GeV.}
\label{fig:4}
\end{figure}

As it is seen from Fig.~\ref{fig:3}, $F_{t}(t)$ for $m_{\sigma }=0.76$ GeV is,
in the physical region of the reaction $\gamma p\rightarrow K^{+}\Lambda$,
rather close to the phenomenological FF down to  --1.5 GeV$^{2}$. At positive
$t$ the FF calculated in the present model keeps increasing and at $t=(\mu_{K}+
m_{\sigma })^{2}$ develops a cusp (not shown explicitly). The latter
corresponds to the physical state of a kaon and a $\sigma$-meson in the
$t$-channel ($p\bar{\Lambda} \rightarrow K^{+}$). The differences between
$F_{t}(t)$ and $F_{t}^{(ph)}(t)$ at $t>0$ may not be considered very important,
because this kinematical region is never reached in the $\gamma p\rightarrow
K^{+}\Lambda $ reaction.

The $s$-channel FF is plotted in Fig.~\ref{fig:3} up to $s\approx 5$ GeV$^{2}$
corresponding to photon lab energies $k_{lab}$ of about 2 GeV. Also here there
is a sharp cusp at $s=(M_{N}+ m_{\sigma })^{2}$ that comes from the
intermediate proton-sigma state depicted in diagrams 2 and 3 in
Fig.~\ref{fig:2}a. However, contrary to the $t$-channel, the difference between
$F_{s}(s)$ and the phenomenological FF $F_{s}^{(ph)}(s)$ shows up in the
physical region at $s\geq (M_{\Lambda}+ \mu_{K})^{2}$. For example, at $s=3$
GeV$^{2}$, corresponding to $k_{lab}\approx $ 1.1 GeV, the FF $F_{s}(s)$ is
larger than $F_{s}^{(ph)}(s)$ by a factor 30. This calculation shows that in
microscopic models the FF's have much richer structure than the commonly used
phenomenological parameterizations.

Fig.~\ref{fig:4} shows the correction to the Born amplitude calculated in the
present model, and in the DW approach, Eq.~(\ref{eq:14}), with FF's from
Eq.~(\ref {eq:21}). Ohta's recipe gives $\Delta A_{2}=0$. It is seen that loop
corrections increase the Born amplitude $A_{2}^{Born}$, in contrast with
the DW prediction. This result depends on the sigma mass; the calculation is
performed with $m_{\sigma }=0.76$ GeV and for the larger mass, $m_{\sigma }= $
$1$ GeV, the correction is less.

\subsection{\label{subsec:Cancellation} Cancellation of loop corrections at large
$m_{\protect\sigma}$}

With increasing $\sigma $-meson mass an interesting effect occurs: the total
correction $\Delta A_{2}$ tends to zero in the loop calculation, implying a
cancellation between the vertex corrections $\mathcal{L}_{[3]}^{\mu}$ and the
4-point loop diagrams $\mathcal{L}_{[4]}^{\mu }$. This cancellation becomes
more complete if $m_{\sigma }\gg M_{N}$ and $ A_{2}\rightarrow A_{2}^{Born}$.
In other words, Ohta's prescription discussed in sect.~\ref{subsec:Vertex} is
reproduced.

There are two ways to understand this effect.  One is by realizing that in
general loop corrections to a strong vertex depend on the four-momenta of all
particles involved. Minimal substitution in such a complicated function
introduces many ambiguities as, for example, was discussed in
Ref.~\cite{Kon00}. In general, gauge-invariant tensor-structures beyond the
minimal substitution can be constructed through a combination of at least two
four-vectors. For these four-vectors one may take the momenta of any particle
involved and/or $\gamma$-matrices when dealing with fermions. The resulting
terms may contribute to any of the tensors $\mathcal{M}^{\mu}_i$ with
$i=1,2,3,4$. Therefore the minimal substitution in the most general
$K^{+}p\Lambda$ vertex does not lead to unique results, even for $A_2 (s,t)$.

If the vertex, however, depends on a single momentum $p^\mu$, then the
procedure of minimal substitution gives unambiguous result for the electric
amplitude. Possible terms beyond minimal substitution in such a vertex are
expressed via $p^\mu$ and  $\gamma^\mu$ solely. Any gauge-invariant structures
built on these two four-vectors will not contribute to $\mathcal{M}^{\mu}_2$,
which the present discussion is focussed on, because the latter tensor involves
two independent momenta $p^{\mu}$ and $p^{\prime \mu}$ (see Eq.(\ref{eq:3})).
In this case any procedure to restore GI should thus yield identical results
for the $A_{2}(s,t)$ amplitude.

\begin{figure}[tbp]
\includegraphics[width=8cm]{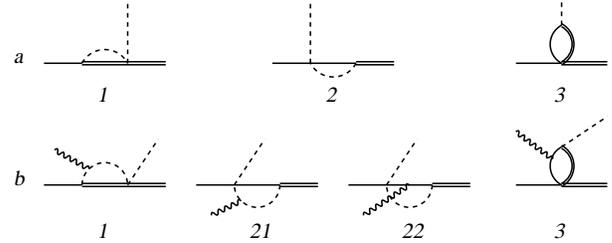}
\caption[A2]{Diagrams describing loop corrections to the 3-point vertex (a) and
4-point vertex (b), in which the $\protect\sigma$-propagator has been contracted to a
point. Notations are the same as in Figs.~\ref{fig:1} and \ref{fig:2}. }
\label{fig:5}
\end{figure}

The diagrams drawn in Fig.~\ref{fig:5}, which correspond to those
of Fig.~\ref{fig:2} in the limit $m_{\sigma }\rightarrow
\infty $, help to understand the situation. In this limit
the propagator of the $\sigma $-meson becomes momentum-independent
and ``shrinks''  to a point-like  interaction, resulting in
effective 4-point vertices $\Lambda
\Lambda K^{+}K^{-}$, $ppK^{+}K^{-}$ and $pp\Lambda \Lambda$.
Simple analysis shows that, for example, diagram  1 in Fig.~\ref{fig:5}a
depends exclusively on the nucleon momentum $p$ and does not depend on the
kaon ($q$) and lambda ($p^{\prime}$) momenta. Therefore this diagram can
generate FF's depending on $s$ only, and cannot lead to any $u$ or $t$
dependencies. Similar arguments apply to other diagrams in Fig.~\ref{fig:5}a.
Any GI restoring procedure therefore gives result which coincides with that of
Ohta~\cite{Oht89}, leading to complete cancellation between the vertex
corrections and the contact terms.
Of course this conclusion is
valid only for the convection current related to $A_2$ and different procedures
may give different results for the magnetic contributions associated with
$A_1$, $A_3$, and $A_4$.

Another way to see the cancellation between the two contributions is to phrase
the problem in term of self-energy corrections. In the limit $m_{\sigma}
\rightarrow \infty$ the vertex correction $\tilde\Gamma_{1}$ depends on $s=p^2$ and
$\vslash{p}$ only and can be rewritten in terms of an irreducible vertex (which
is point-like in this limit) and a self-energy correction, $\tilde{\Gamma}_{1}
=\frac{g}{2f_{K}} \sqrt{3} \sin (\alpha -\Delta \alpha )
\gamma_{5}\Sigma_{N}(p) \approx  \frac{g}{2f_{K}}\gamma_{5}\Sigma_{N}(p)$.
Since the vertex is normalized at the on-shell point $p^2 = M_{N}^2$, the
correction vanishes there, which implies that the self-energy also vanishes on
shell, {\it i.e.} $\Sigma_{N}(p)u(p)=0$. The 4-point term, shown in diagram 1
in Fig.~\ref{fig:5}b, is proportional to $\tilde{\Gamma}^{\mu }(p+k,p)$, that
corresponds to a photon coupling to the vertex correction $\tilde\Gamma_{1}$.
It can be shown algebraically that
\begin{eqnarray}
k_{\mu }\ \tilde{\Gamma}^{\mu}(p+k,p)u(p)
 &=& [\Sigma_{N}(p)-\Sigma_{N}(p+k)]u(p) \nonumber \\
 &=& -\Sigma _{N}(p+k)u(p).
\label{eq:22}
\end{eqnarray}
One may also argue that since the photon is coupled to all charged particles in
the loop the vertex should obey the Ward-Takahashi identity~\cite{Itz} which
reduces to Eq.~(\ref{eq:22}) for the vertex correction.

The total correction to the Born amplitude can be written as
\begin{eqnarray}
 &\frac{eg}{2f_{K}}& \bar{u}(p^{\prime })\gamma_{5}
 \big[ \Sigma_N(p+k) S_0(p+k) \gamma ^\mu +\tilde{\Gamma}^\mu(p+k,p) \big]u(p)
 \nonumber \\ &\equiv&
 \frac{eg}{2f_{K}} \bar{u}(p^{\prime })\gamma _{5}J^{\mu }u(p),
\label{eq:23}
\end{eqnarray}
where $S_{0}(p+k)= (\vslash{p}+ \vslash{k}-M_{N})^{-1}$ is the free proton
propagator. It is now straightforward to show that the current $J^{\mu}u(p)$ is
\ a) purely transverse, $\ k_{\mu }J^{\mu }u(p)=0$, and \ b) independent of the
momenta $p^{\prime }$ of the $\Lambda$ and $q$ of the kaon. The first condition
implies that the matrix element in Eq.~(\ref{eq:23}) can be expressed in terms
of the four Lorentz spin-tensors $\mathcal{M}^\mu_i$, and the second implies
that only $i=1,3$ are allowed. The contribution to the convection current
therefore vanishes, {\it i.e.}  $\Delta A_{2}=0$.

This result is general, although the arguments for the different diagrams in
Fig.~\ref{fig:5} differ in detail. Diagram 2 in Fig.~\ref{fig:5}a, for example,
describes the vertex correction $\tilde{\Gamma}_{2}$. This correction is
proportional to an effective self-energy $\Sigma _{\Lambda }$ of the lambda.
Since the final lambda is on its mass shell this self-energy vanishes. The
corresponding amplitude, proportional to $\tilde{\Gamma}_{2}$, vanishes as
well. The 4-point terms are given by the diagrams 21 and 22 in
Fig.~\ref{fig:5}b. Each of them is not zero, however they have opposite signs
and cancel each other in $A_{2}$ when $m_{\sigma }\rightarrow \infty$. More
formally, this result follows from the fact that diagrams 21 and 22 describe a
correction $\tilde{\Gamma}^{\mu }(p^{\prime}, p^{\prime}-k)$ to the EM vertex
of the neutral lambda-hyperon, which is transverse, $k_{\mu} \cdot
\tilde{\Gamma}^{\mu} (p^{\prime}, p^{\prime}-k) =0$, and independent of the
momenta $p$ of the proton and $q$ of the kaon. Therefore these diagrams
contribute to the tensors $\mathcal{M}^\mu_i$ with $i=1,4$ only, \ and  $\Delta
A_{2}=0$.

The situation with the corrections described by diagrams 3 in
Fig.~\ref{fig:5} is basically similar to diagrams 1. We only have to take
into account that, since the couplings $g_{\sigma pp}$ and $g_{\sigma
\Lambda \Lambda }$ are independent of $m_{\sigma }$, both 3- and 4-point
corrections diminish if $ m_{\sigma }\rightarrow \infty$. In order to observe
effect of the cancellation,  the diagrams have to be kept finite. One can
assume, somewhat artificially, that the product $g_{\sigma pp}g_{\sigma \Lambda
\Lambda }$ rises linearly with $m_{\sigma }$.
A reasoning similar to that given above leads to the conclusion that $\Delta
A_{2}=0$. Moreover, because the kaon is a spinless particle, the loop
corrections do not contribute to the amplitudes $A_{1},A_{3}$ and $A_{4}$ as
well.

\begin{figure}[tbp]
\includegraphics[width=6cm]{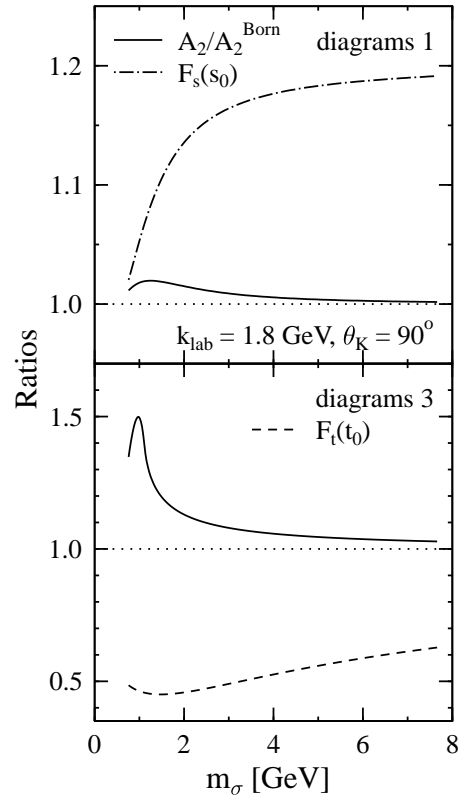}
\caption[Ratio]{Ratio $A_{2} /A_{2}^{Born}$ and form factors in
the $s$- and $t$-channels as functions of $\protect\sigma$-meson mass. Upper
panel: calculation including only diagrams 1 in Fig.~\protect\ref{fig:2}, lower
panel: calculation including only diagrams 3 in Fig.~\protect\ref{fig:2}.
For diagrams 3 the coupling $g_{\protect\sigma pp}g_{\protect\sigma \protect\Lambda
\protect\Lambda }$ is multiplied by $m_{\protect\sigma}/0.76$ GeV.
The invariant energy and momentum transfer corresponding to this kinematics
are $s_0 = 4.26$ GeV$^2$ and $t_0 = - 1.05$ GeV$^2$ respectively.}
\label{fig:6}
\end{figure}

The above considerations are supported by numerical calculations (see
Fig.~\ref{fig:6}). It is seen from the upper part of Fig.~\ref{fig:6} that,
when only the diagrams 1 in Fig.~\ref{fig:2} are included, the FF $F_s(s)$
differs considerably from unity. This means that the correction to the 3-point
vertex does not vanish in the limit $m_{\sigma} \to \infty$ and yields the
$s$-dependence of the vertex. In this limit the diagram should not contribute
to the $t$-channel FF and indeed one finds that $F_t(t) \approx 1$ (not shown).
At the same time the amplitude $A_2$ approaches the Born term at large
$\sigma$-mass indicating the cancellation of all corrections as discussed
above. It appears also that the corrections are small even at moderate
$m_\sigma$. This is just a consequence of the fact that the coupling constants
which enter this diagram are small.

If only the diagrams 3 in Fig.~\ref{fig:2} are switched on, the $t$-dependence
of the $K^+ p \Lambda$ vertex is generated  at all values of $m_\sigma$
(see the lower part of Fig.~\ref{fig:6}). Nonetheless the correction to
the Born amplitude decreases fast and is practically negligible at $m_\sigma$
about 3 - 4 GeV. It is also interesting to note that at $m_\sigma \approx 1$
GeV the total correction is maximal. This reflects non-regular behavior of the
4-point loop diagram as a function of invariant energy (cusp structure due to
nucleon-sigma intermediate state).

The arguments presented above also justify our neglect of the proper
self-energy insertions and corrections to the electromagnetic vertices,
mentioned in the beginning of sect.~\ref{subsec:Loops}, in the calculation of
the electric amplitude $A_2$.

We can conclude that in a microscopic model, where the 3-point vertex has a
particular structure which generates off-shell dependence only on a single
momentum (corresponding to the limit where one of the particles in the loops
is very heavy), different corrections to the Born amplitude cancel
and the procedure of Ohta~\cite{Oht89} is justified.

In more realistic models, in which light mesons
are present in the loops, the corrections do not cancel and the
DW procedure may be more appropriate.  To test this assumption we
calculated $\Delta A_2^{DW}$ in Eq.(\ref{eq:14})
with the microscopic FF's shown in  Fig.~\ref{fig:3},
instead of phenomenological FF's used in the original formulation
(one may argue however that only the phenomenological FF's are
to be used in the DW approach).
It turns out that at small $m_{\sigma}$ the correction
$\Delta A_2^{DW}$ is of the same sign and similar magnitude
as $\Delta A_2^{loops}$, however results disagree in the conditions
where one is close to a threshold for the production of
the physical particles included in the loop diagrams.
While  $\Delta A_2^{loops}$ is a flat and smooth background
(see, for example, the dashed curve in Fig.~\ref{fig:4}),
the DW calculation shows an irregular behavior reflecting the
cusps in $F_s (s)$.

The above observations may support the picture in which the DW approach
is applicable in the regime where loop corrections arise due
to the light intermediate particles,
while the method of Ohta is applicable in the other extreme.

The precise magnitude of the loop corrections depends of course
on the detailed structure of the model. In the present semi-realistic
calculation, where pions and sigma-hyperons are not taken into account, the
corrections tend to enhance the Born amplitude. This situation may
however be reversed if all possible intermediate states are included.

Finally we should mention that the diagrams similar to those in
Fig.~\ref{fig:2}, where $f_0$-meson replaces the $\sigma$, have also been
included in the calculation. The effect of these turns out to be very small and
can be discarded.

\section{\label{sec:Conclusions}Conclusions}

In the present work we have explicitly calculated  loop corrections to the
strong  $K^{+}p\Lambda$ vertex  and the additional 4-point amplitude in an
effective Lagrangian model for photo-induced $K^{+} \Lambda $ production off
the proton. The main focus has been on the scalar amplitude $A_2 (s,t)$,
associated with the electric contribution in a gauge-invariant approach.

The calculation shows that there is a strong cancellation in the amplitude
$A_2(s,t)$ between 3-point loop corrections, often parameterized via form
factors in a phenomenological approach, and 4-point loop corrections, often
written as contact terms restoring gauge invariance. This cancellation becomes
complete if one of the intermediate particles in the diagrams becomes
infinitely heavy. This shows that the result of the minimal-substitution method
of Ohta~\cite{Oht89} can be understood in the microscopic picture as a
particular limiting case.

In a more realistic case the cancellation is not complete. Only part of the
loop corrections can be absorbed in an appropriately chosen form factors and
contact terms, constructed using the minimal substitution. The remaining part
can be accounted through 4-point contact terms which are gauge invariant
and free of the poles, as was done in the work of DW~\cite{Dav01,Dav01-2}.

The calculation also indicates that the form factors, usually taken in
phenomenological approaches, may not be realistic. In a microscopic model the
form factors are necessarily complex, with cusp structures in the real part
reflecting the possibility that the intermediate states become physical
particles for certain kinematical conditions. Even though we have focussed our
attention in this calculation on the electric current, we expect that the
conclusion about the complex structure of the form factors is more general and
applies to any vertex, which accounts for loop corrections not explicitly
included in the present model. Non-trivial structures of the form factors
extracted from a microscopic calculation were also observed in
Ref.~\cite{Kon01}. However the cancellation of different contributions which
was observed in the calculation of the $A_2$ term may, due to the current
conservation, be a peculiarity of the electric amplitude.

In the present calculation only the scalar mesons, proton, kaon and
lambda-hyperon have been included in the loop diagrams. In particular
contributions involving the $\Sigma$-baryon and the pion have not been
considered. Qualitatively one expects these particles to give similar
contribution (in absolute magnitude and structure, but sign may be different)
to what has been calculated in this work. Clearly a more complete calculation,
where all possible intermediate states are taken into account, is needed to
adequately describe the non-resonant part of the $\gamma p \to K^{+}
\Lambda$ amplitude. The knowledge of the latter is imperative to separate
the resonances from the background contribution in strangeness photoproduction.


\begin{acknowledgments}
Part of this work was performed as part of the research program
of the Stichting voor Fundamenteel Onderzoek der Materie
(FOM) with financial support from the Nederlandse Organisatie
voor Wetenschappelijk Onderzoek (NWO).
One of the authors (A.Yu.K.) acknowledges a grant from
the NWO. He would also like to thank the staff
of the Kernfysisch Versneller Instituut for the kind hospitality.
We thank Rob Timmermans for helpful discussions.
\end{acknowledgments}


\bigskip \appendix

\section{\label{App:A} Evaluation of loop integrals for the $pK^{+}\Lambda $
vertex and $\protect\gamma p\rightarrow K^{+}\Lambda $ amplitude}

The one-loop vertex corrections, corresponding to the diagrams on
Fig.~\ref{fig:2}a can be written as
\begin{eqnarray}
\tilde{\Gamma}_{1} &=& gg_{\sigma \Lambda \Lambda} g_{\sigma K^+K^-}
 \frac{i}{(2\pi )^{4}} \nonumber \\
 &&\int
 \frac{\gamma_{5}(2M_{\Lambda }+L\hspace{-0.5em}/)\text{d}^{4}L}
 {[(p^{\prime }-L)^{2}-M_\Lambda^{2})][(L+q)^{2}-\mu_K^2](L^2-m_\sigma^2)},
   \nonumber \\
\tilde{\Gamma}_{2} &=&gg_{\sigma pp}g_{\sigma K^{+}K^{-}}\frac{i}{(2\pi
)^{4}} \nonumber \\
 &&\int \frac{\gamma _{5}(M_{N}+\vslash{p}-L\hspace{-0.5em}/)\text{
d}^{4}L(2\pi )^{-4}}{[(p-L)^{2}-M_{N}^{2})][(L-q)^{2}-\mu
_{K}^{2}](L^{2}-m_{\sigma }^{2})},  \nonumber \\
\tilde{\Gamma}_3 &=&gg_{\sigma pp}g_{\sigma \Lambda \Lambda }
 \frac{i}{(2\pi )^{4}} \label{eq:A1} \\
 &&\int \frac{\gamma _{5}(2M_{\Lambda }+L\hspace{-0.5em}/)(M_{N}+p%
\hspace{-0.5em}/-L\hspace{-0.5em}/)\text{d}^{4}L(2\pi )^{-4}}{%
[(p-L)^{2}-M_{N}^{2})][(p^{\prime }-L)^{2}-M_{\Lambda }^{2}](L^{2}-m_{\sigma
}^{2})}, \nonumber
\end{eqnarray}
where we explicitly used Dirac equation for the final lambda, however did not
assume the initial proton and the kaon to be on their the mass shells, {\it
i.e.} $p^{2}=s\neq M_{N}^{2}$ and $q^{2}=t\neq \mu _{K}^{2}$. Using the Feynman
parametrization and integrating over the loop momentum $L$ we obtain
\begin{eqnarray}
\tilde{\Gamma}_{1} &=&2gC_{1}\gamma_{5}\int_{0}^{1}\text{d}x\int_{0}^{x}
\text{d}y \frac{M_{\Lambda }(2-x)-\vslash{p}(x-y)}{\Delta_{1}(x,y)}
,  \nonumber \\
\tilde{\Gamma}_{2} &=&2gC_{2}\gamma_{5}\int_{0}^{1}\text{d}x\int_{0}^{x}
\text{d}y \frac{M_{N}-M_{\Lambda }(x-y)+\vslash{p}(1-x)}{\Delta
_{2}(x,y)},  \nonumber \\
\tilde{\Gamma}_{3} &=& 2gC_{3}\gamma_{5}\int_{0}^{1}\text{d}x\int_{0}^{x}
\text{d}y \Big\{ 2N_\epsilon-1-2\ln \Delta_{3}(x,y) \nonumber \\
 &+& \big[ M_{N}M_\Lambda (2-x+y)+M_{\Lambda }^{2}(2-x)(x-y)  \nonumber \\
 &+& \mu _{K}^{2}y(x-y)+sy(1-x)+\vslash{p}(M_{\Lambda }(2-x-y)  \nonumber \\
 &+& M_{N}y) \big] \frac{1}{\Delta _{3}(x,y)} \Big\},  \label{eq:A2}
\end{eqnarray}
with $C_{1}={g_{\sigma \Lambda \Lambda}g_{\sigma K^+ K^-} / 32\pi^{2}}$,
$C_{2}=g_{\sigma pp}g_{\sigma K^{+}K^{-}}/ 32\pi^{2}$, $C_{3}=g_{\sigma
\Lambda \Lambda }g_{\sigma pp}/32\pi^{2}$, and
\begin{eqnarray}
\Delta _{1}(x,y) &=& M_{\Lambda }^{2}xy-sy(x-y)\nonumber \\
 &+& [\mu_{K}^{2}-t(1-x)](x-y)
 + m_{\sigma }^{2}(1-x) -i0,  \nonumber \\
\Delta _{2}(x,y) &=&y[M_{N}^{2}-s(1-x)]-M_{\Lambda }^{2}y(x-y) \nonumber \\
 &+& [\mu_{K}^{2}-t(1-x)](x-y)+m_{\sigma }^{2}(1-x) -i0,  \nonumber \\
\Delta _{3}(x,y) &=&y[M_{N}^{2}-s(1-x)]+M_{\Lambda}^{2}x(x-y) \nonumber \\
 &-& ty(x-y)+m_{\sigma }^{2}(1-x) -i0.  \label{eq:A3}
\end{eqnarray}

While $\tilde{\Gamma}_{1}$ and $\tilde{\Gamma}_{2}$ are convergent,
$\tilde{\Gamma}_{3}$ has a divergent piece which in the
dimensional-regularization method is expressed via the constant $N_{\epsilon
}=2/\epsilon -\gamma _{E}+\ln 4\pi$, where $\epsilon \equiv 4-D$ and $D$ is the
space-time dimension. In order to normalize the vertex we add a counter term of
the form $\delta g\,\gamma_{5}$ to the loop corrections. The coefficient
$\delta g$ can be fixed by requiring $g$ to be the physical coupling constant
$g_{K^{+}p\Lambda}$. This condition automatically makes the vertex finite and
properly normalized to $g\gamma_{5}$ for on-mass-shell particles. Subsequently
the off-shell  vertex is calculated from
\begin{eqnarray}
\bar{u}(p^{\prime }) \Gamma (p^{\prime },p;q)=\bar{u}(p^{\prime }) \big[
 g\gamma_{5} +\tilde{\Gamma}_{1}+ \tilde{\Gamma}_{2}+ \tilde{\Gamma}_{3}
  \nonumber \\
 -(\tilde{\Gamma}_{1}+ \tilde{\Gamma}_{2}+ \tilde{\Gamma}_{3})_{\
p^{2}=M_{N}^{2},\ q^{2}=\mu _{K}^{2}} \big] \;.  \label{eq:A4}
\end{eqnarray}
All FF's defined in Eqs.~(\ref{eq:5}) and (\ref{eq:6}) of
sect.~\ref{subsec:Vertex} can be obtained from Eq.~(\ref{eq:A4}). In
particular, the $s$- and $t$-channel FF's are normalized as follows
\begin{equation}
F_{s}(M_{N}^{2})=F_{t}(\mu _{K}^{2})=1 \;.  \label{eq:A4-1}
\end{equation}
There is an essential difference between the $s-$ and $t-$channels.  In the
physical region of the $\gamma p\rightarrow K^{+}\Lambda $ reaction, where
$t<0$, the function $F_{t}(t)$ is real. In this case the dominators in
Eqs.~(\ref{eq:A2}) do not vanish and the calculation of the two-fold integrals
is straightforward. In the $s-$channel however the FF has an imaginary part
which develops at $s\geq (M_{\Lambda }+\mu _{K})^{2}$ for the 1$^{st}$ diagram
in Fig.~\ref{fig:2}a and at $s\geq (M_{N}+m_{\sigma })^{2}$ for the 2$^{nd}$
and 3$^{rd}$ diagrams. Calculation of the corresponding integrals requires care
and we apply the methods developed by 't Hooft and Veltman in
Ref.~\cite{tHo79}. In particular, one integration in Eqs.~(\ref{eq:A2}) can be
performed which ensures that the remaining integrals are numerically stable.

\begin{widetext}

The 4-point loop contributions are shown in Fig.~\ref{fig:2}b. We introduce the
notation $\mathcal{L}_{[4]}^{\mu}= \mathcal{L}_{[4],1}^{\mu } +
\mathcal{L}_{[4],21}^{\mu }+ \mathcal{L}_{[4],22}^{\mu }+
\mathcal{L}_{[4],3}^{\mu }$ for the four diagrams in Fig.~\ref{fig:2}b. For
example, for $\mathcal{L}_{[4],1}^{\mu }$ we have
\begin{eqnarray}
&& \mathcal{L}_{[4],1}^{\mu } =egg_{\sigma \Lambda \Lambda }g_{\sigma
K^{+}K^{-}}\frac{i}{(2\pi )^{4}} 
 \int \frac{\gamma _{5}(2M_{\Lambda }+L
\hspace{-0.5em}/)[2(L^{\mu }+q^{\mu })-k^{\mu }]\text{d}^{4}L}{[(p^{\prime
}-L)^{2}-M_{\Lambda }^{2})][(L+q)^{2}-\mu _{K}^{2}][(L+k-q)^{2}-\mu
_{K}^{2}](L^{2}-m_{\sigma }^{2})} \;,  \label{eq.A5}
\end{eqnarray}
and similarly for $\mathcal{L}_{[4],21}^{\mu }, \mathcal{L}_{[4],22}^{\mu }$
and $\mathcal{L}_{[4],3}^{\mu }$ (the baryon spinors are omitted for brevity).
One can explicitly check that these amplitudes satisfy the relation
\begin{eqnarray}
&& k\cdot ( \mathcal{L}_{[4]1} + \mathcal{L}_{[4]21} + \mathcal{L}_{[4]22} +
\mathcal{L}_{[4],3}) = - k\cdot \mathcal{T}_{El} 
 =-e\bar{u}(p^\prime) [\Gamma(p^\prime,p+k;q) -\Gamma(p^\prime,p;q-k)]u(p) \;,
\label{eq:A6}
\end{eqnarray}
which guarantees GI of the total amplitude.

We perform integration over $L,$ as well as over one of the Feynman parameters.
In order to obtain contribution of these loops to the coefficient $H_{t}(s,t)$
in Eq.~(\ref{eq:14}) one has to project the tensor structure $\bar{u}(p^{\prime
})\gamma _{5}q^{\mu }u(p)$ out of the final result. Any of the 4-point
integrals is proportional to the tensors: $\gamma_{5} \gamma^\mu, \
\gamma_{5}\gamma^\mu \vslash{k}, \
\gamma_{5} p^{\mu} \vslash{k}, \
\gamma_{5} q^{\mu} \vslash{k},
\ \gamma_{5} p^{\mu}$ and $ \gamma_{5} q^{ \mu}$,
and of course terms $\propto  \gamma_{5} k^{\mu}, \gamma_{5} k^{\mu}\vslash{k}$
can be dropped.
It suffices for our purposes to select only those proportional to $\gamma_{5}q^{\mu}$.
In this way we arrive at the integrals with the following generic
structure
\begin{eqnarray}
&&H_{t}(s,t) =\text{const}_{1}\int_{0}^{1}\text{d}x\int_{0}^{x}y\text{d}y\frac{%
\{1,x,x^{2},y,y^{2},xy\}}{A(A+By)} 
 + \text{const}_{2}\int_{0}^{1}\text{d}x\ln
\frac{M_{\Lambda }^{2}(1-x)+M_{N}^{2}x-tx(1-x)}{M_{\Lambda
}^{2}(1-x)+M_{N}^{2}x-\mu _{K}^{2}x(1-x)},  \label{eq:A7}
\end{eqnarray}
\end{widetext}
with the polynomials
\begin{eqnarray}
A &=&ax^{2}+by^{2}+cxy+dx+ey+f-i0,  \nonumber \\
B &=&hx+jy+k.  \label{eq:A8}
\end{eqnarray}
The coefficients $a,b,c,d,e,f,h,j,k$ depend on a particular diagram and
are expressed in terms of the masses of external and internal particles, and
the Mandelstam variables $s,u$ and $t$. The second integral in
Eq.~(\ref{eq:A7}) appears only in $\mathcal{L}_{[4],3}^{\mu }$. One integration
can further be done using the methods of Ref.~\cite{tHo79}. The obtained
coefficients $H_{t}(s,t)$ for the diagrams in Fig.~\ref{fig:2}b
are calculated numerically. In the numerical calculation double precision
and a large number of mesh points are used to get accurate results.


\end{document}